\documentclass[journal]{IEEEtran}
\ifCLASSINFOpdf
  \usepackage[pdftex]{graphicx}
\else
\fi
\usepackage{amsmath}
\usepackage{amssymb}
\ifCLASSOPTIONcompsoc
 \usepackage[caption=false,font=normalsize,labelfont=sf,textfont=sf]{subfig}
\else
 \usepackage[caption=false,font=footnotesize]{subfig}
\fi

\usepackage{siunitx}
\begin{document}
%
\title{On the Influence of Smoothness Constraints in Computed Tomography Motion Compensation}

\author{Mareike~Thies,
        Fabian~Wagner,
        Noah~Maul,
        Siyuan~Mei,
        Mingxuan~Gu,
        Laura~Pfaff,
        Nastassia~Vysotskaya,
        Haijun~Yu,
        and~Andreas~Maier%
\thanks{All authors are with the Pattern Recognition Lab, Friedrich-Alexander-Universit\"at Erlangen-N\"urnberg, Erlangen, Germany, Correspondence e-mail: mareike.thies@fau.de.}}

\maketitle

\begin{abstract}
Computed tomography (CT) relies on precise patient immobilization during image acquisition. Nevertheless, motion artifacts in the reconstructed images can persist. Motion compensation methods aim to correct such artifacts post-acquisition, often incorporating temporal smoothness constraints on the estimated motion patterns. This study analyzes the influence of a spline-based motion model within an existing rigid motion compensation algorithm for cone-beam CT on the recoverable motion frequencies. Results demonstrate that the choice of motion model crucially influences recoverable frequencies. The optimization-based motion compensation algorithm is able to accurately fit the spline nodes for frequencies almost up to the node-dependent theoretical limit according to the Nyquist-Shannon theorem. Notably, a higher node count does not compromise reconstruction performance for slow motion patterns, but can extend the range of recoverable high frequencies for the investigated algorithm. Eventually, the optimal motion model is dependent on the imaged anatomy, clinical use case, and scanning protocol and should be tailored carefully to the expected motion frequency spectrum to ensure accurate motion compensation. 
\end{abstract}


\section{Introduction}
%
%
%
%
\thispagestyle{empty} 
\IEEEPARstart{F}{or} computed tomography (CT), it is crucial that the patient does not move throughout the acquisition of the scan. Patient motion leads to a mismatch between the measured sinogram and the geometry settings assumed during reconstruction which results in artifacts in the reconstructed image. Motion compensation approaches consequently aim to eliminate these artifacts in a post-processing step. 

Whereas some methods act only on the reconstructed, motion-affected image \cite{Ko2021}, many works explicitly estimate the underlying motion curves and perform a motion-compensated reconstruction given an updated geometry \cite{Aichert2015,Berger2016,Sisniega2017,Preuhs2020}. Explicit modelling of the motion patterns ensures that the motion-compensated reconstruction is consistent with the measured sinogram and yields interpretable solutions where estimated motion curves can be inspected for plausibility. 

Most approaches, which estimate motion curves explicitly, enforce a type of smoothness constraint on the motion patterns over time \cite{Berger2016,Sisniega2017,Preuhs2020}. Usually, this is achieved by either adding a regularization term to the objective function which implicitly punishes non-smooth motion patterns, or explicitly approximating the motion patterns with a smooth function. Using such smoothness constraints makes sense intuitively, since realistic motion patterns are temporally correlated especially considering the high frame rate at which projection images are acquired in modern scanners. Moreover, it can simplify the optimization problem of motion estimation by smoothing the loss landscape or by reducing the number of free parameters which need to be estimated. However, the influence of these modelling decisions on the type of motion patterns which can be recovered by the particular method is often not investigated. 

In this work, we perform a frequency analysis for our previously published motion compensation algorithm \cite{Thies2024} which features an explicit spline-based representation of the motion patterns. Specifically, we formulate the following research questions:
\begin{itemize}
    \item Which frequencies in the patient's motion patterns can be reliably recovered by the algorithm?
    \item How do these frequencies relate to the choice of motion model parameterizing the recovering motion patterns?
    \item Can we loosen the smoothness constraint to increase the range of recovering motion patterns that can be fitted? 
\end{itemize}

\section{Methods}
\begin{figure}[t!]
    \centering
    \includegraphics[width=\columnwidth]{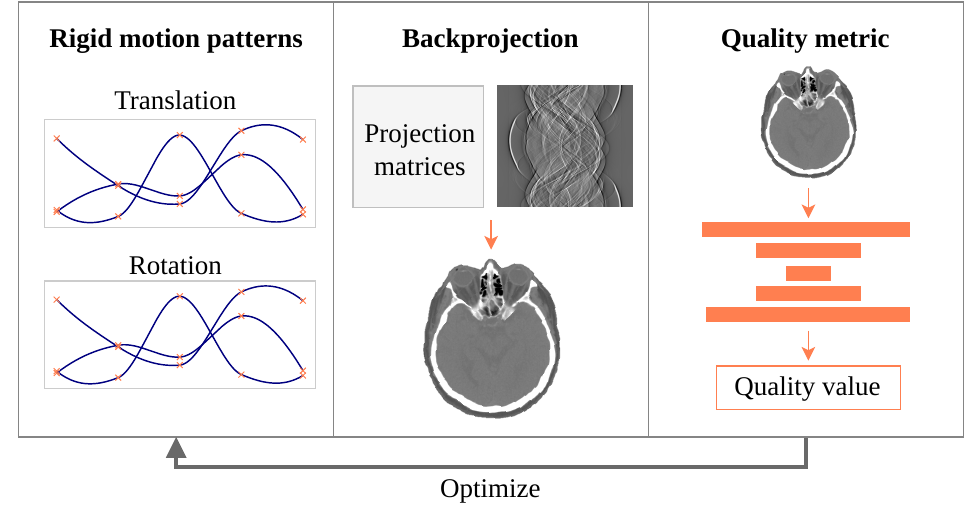}
    \caption{High level overview of the motion compensation algorithm studied in this paper.}
    \label{fig:moco_overview}
\end{figure}
\subsection{Motion Compensation Algorithm}
This work builds up on our previously developed rigid motion compensation algorithm for cone-beam CT systems \cite{Thies2024}. It estimates motion patterns for the six rigid motion parameters ($t_x$, $t_y$, $t_z$, $r_x$, $r_y$, $r_z$) over the scan range. Here, $t_{\Box}$ represents translation along one axis and $r_\Box$ represents rotation around one axis (Euler angles). The estimation of the motion patterns is performed by formulating an optimization problem. The target function is inspired by the autofocus principle where intermediate reconstructions are computed given the current estimate of the motion parameters. A quality metric quantifies the amount of left-over artifacts in the intermediate reconstruction and guides the optimization toward motion patterns yielding reconstructions with maximal quality. For the intermediate reconstruction, we use a backprojector for which an analytic Jacobian is computed. The Jacobian connects the space of intermediate reconstructions and the space of cone-beam geometries (parameterized by projection matrices) in a differentiable manner \cite{Thies2023}. Based on this, gradients of the quality metric to the free parameters of the recovering motion patterns are obtained, and gradient-based optimization is applied. The quality metric is a reference-free mapping which grades the quality of the intermediate reconstruction given only the motion-affected volume. It is parameterized by a neural network trained to regress the visual information fidelity (VIF) \cite{Sheikh2006} as a spatially resolved volumetric map from the input reconstruction followed by a simple average resulting in a scalar quality value. A high-level graphical summary of the most important steps of the motion compensation algorithm is presented in Fig.~\ref{fig:moco_overview} and we refer the interested reader to our paper \cite{Thies2024} for a more in-depth description.

\subsection{Spline-Based Motion Model}
\label{sec:spline}
The motion compensation algorithm in \cite{Thies2024} enforces smoothness of the separate rigid motion parameters. This is achieved by approximating the parameter's values over time with smooth Akima splines \cite{Akima1970}. Each spline is parameterized by $N_n$ nodes which are distributed over the full scan range with equidistant spacing and with one node aligning with the first and last projection of the scan, respectively. Given that a scan has $N_p$ projections, the spline-based formulation limits the number of free parameters which need to be fitted during motion estimation from $6N_p$ for the unconstrained setting to $6N_n$ for the setting with smoothness constraint ($N_n < N_p$). 

\subsection{Frequency Analysis}
\label{sec:frequency}
\begin{figure}
    \centering
    \includegraphics[width=\columnwidth]{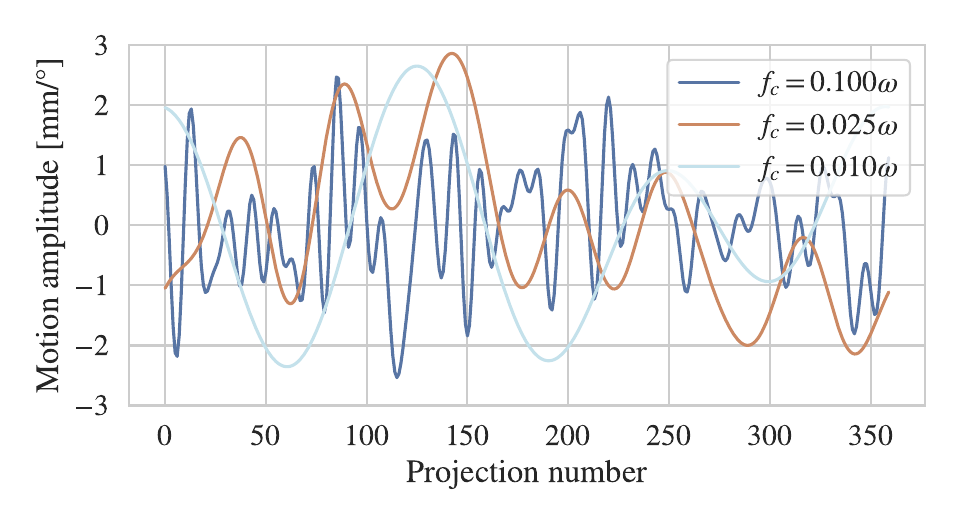}
    \caption{Exemplary motion patterns obtained after limiting a random signal to the cutoff frequency $f_c$. Motion patterns with a low cutoff frequency oscillate slower than those with high cutoff frequencies.}
    \label{fig:motion_examples}
\end{figure}
To analyze the frequency behavior, data is perturbed with random motion signals of different band-limitation in frequency domain. For each of the six rigid parameters, an unconstrained\footnote{In this study, we always refer to unconstrained \textit{observable} patterns, i.e., sampled with the scanner-specific sampling rate.} random motion pattern $x$ is sampled. Subsequently, a Fourier transformation $\mathcal{F}$ is applied to the randomly sampled pattern and all frequencies higher than a certain cutoff frequency $f_c$ are set to zero. After an inverse Fourier transform $\mathcal{F}^{-1}$, the signal is rescaled to a predefined maximal amplitude. The resulting motion pattern $x_c$ is
\begin{equation}
    x_c = r(\mathcal{F}^{-1}(b(\mathcal{F}(x), f_c))) \enspace ,
\end{equation}
where $b(\cdot,\cdot)$ performs the band-limitation of the signal in Fourier domain with cutoff frequency $f_c$ and $r$ rescales the amplitude of the signal. Three examples of band-limited motion patterns along with their corresponding cutoff frequency are shown in Fig.~\ref{fig:motion_examples}. All investigations in this paper are performed independently of the scanner-specific sampling rate $\omega$, measured in $1/\text{s}$, at which projections are acquired. For an entirely random motion pattern without smoothness constraint, the maximal frequency that can be measured given a certain scanner-dependent sampling rate $\omega$ is $f_{max} = 0.5 \omega$, which follows from the fundamental Nyquist-Shannon sampling theorem. Any cutoff frequency $f_c < f_{max}$ smaller than this maximal frequency reduces the highest frequency present in the motion signal. Cutoff frequencies are sampled such that they are equidistant in logarithmic space. $f_c = 0.005 \omega$ is the smallest investigated frequency and $f_c = f_{max}$ is the highest investigated frequency which corresponds to unconstrained motion.      

\section{Data and Experiments}
The experiments are based on the publicly available CQ500 data set \cite{Chilamkurthy2018} which contains reconstructed head CT volumes acquired on different helical multi-detector CT scanners. For the investigations in this paper, we use a subset of ten scans which are reconstructed with a small slice thickness of \qty{0.625}{\milli\meter}. These scans have not been incorporated in the training or development of the quality metric regression network which is part of the motion compensation pipeline. Cone-beam sinograms are obtained from the reconstructed volumes via forward projection using the software \textit{PyroNN} \cite{Syben2019}. We simulate a (potentially motion-perturbed) circular full scan with 360 projections, a source-to-isocenter distance of \qty{785}{\milli\meter}, a source-to-detector distance of \qty{1200}{\milli\meter}, and a detector of shape $500 \times 700$ pixels of isotropic pixel spacing (\qty{0.64}{\milli\meter}). The motion compensation algorithm is run on reconstructed volumes of $128^3$ voxels with an isotropic spacing of \qty{2}{\milli\meter}. For evaluation, we reconstruct the data on grids with $256^3$ voxels and a spacing of \qty{1}{\milli\meter} in all dimensions.  

Motion compensation is performed by executing the following steps: First, a random, band-limited motion signal is sampled as described in section~\ref{sec:frequency} using a maximal amplitude of \qty{5}{\milli\meter} for translation and \ang{5} for rotation. This signal is used to rigidly perturb the ideal circular trajectory and a volume from the data set is forward projected onto this motion-affected trajectory. The motion compensation algorithm is initialized with the ideal circular trajectory. It estimates the perturbing motion pattern as precisely as possible based on the spline-based representation explained in section~\ref{sec:spline}.  

\begin{figure*}
    \centering
    
    \subfloat{\includegraphics[width=\columnwidth]{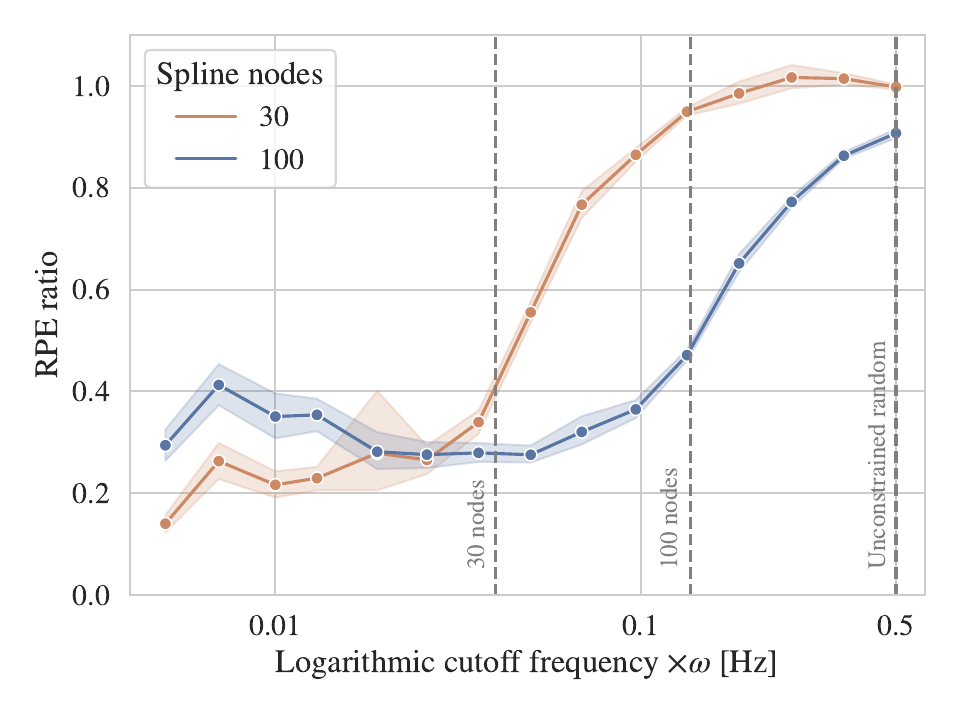}\label{fig:result_rpe}}
    \subfloat{\includegraphics[width=\columnwidth]{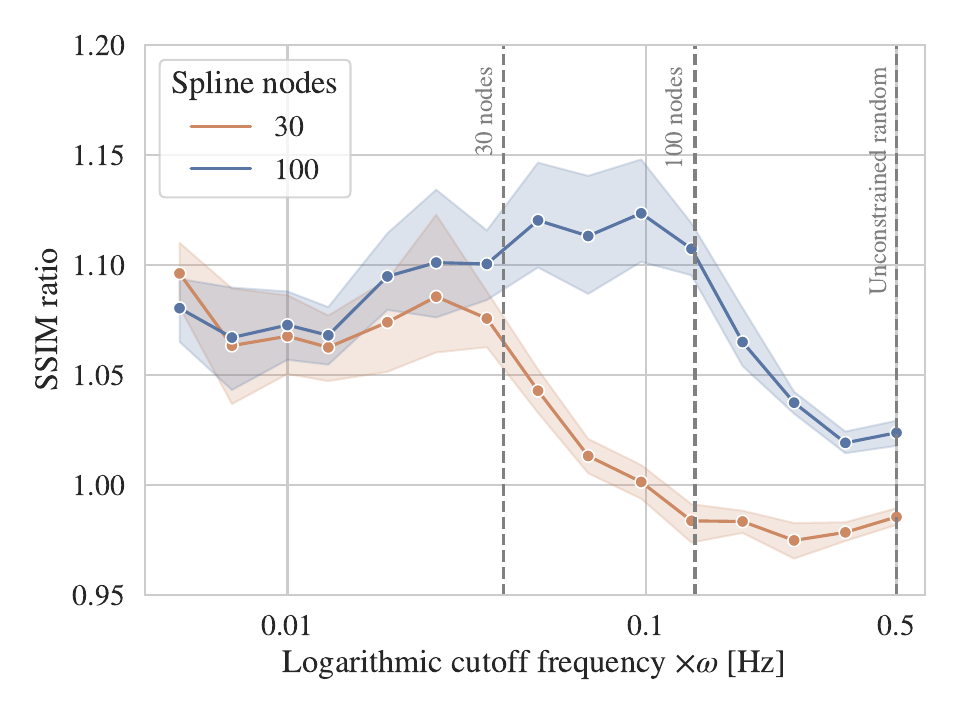}\label{fig:result_ssim}}
    
    \caption{Plot of the change in reprojection error (RPE) (left) and structural similarity index measure (SSIM) (right). Both are displayed as their ratio of after/before motion compensation. While smaller values for RPE and higher values for SSIM are better, respectively, a ratio close to 1 reflects no improvement after motion compensation for either metric. The ratio is drawn as a function of the cutoff frequency $f_c$ in the perturbing motion pattern. Note that the x-axis is scaled logarithmically from small to high frequencies. The maximum measurable frequencies according to the Nyquist-Shannon theorem are indicated by gray dashed lines for both spline types and the unconstrained setting. The two curves show the results obtained with splines with 30 nodes and 100 nodes for the estimated motion signal. \qty{95}{\%} confidence intervals are computed over ten different scans.}
    \label{fig:overall}
\end{figure*}

\section{Results}
The motion compensation algorithm is applied to ten scans. 15 different perturbing motion patterns with varying cutoff frequencies $f_c$ are applied to each sample. Each combination is recovered using 30 and 100 nodes for the spline modelling the recovering signal leading to $10 \cdot 15 \cdot 2 = 300$ runs in total. We compute the reprojection error (RPE) which measures deviations between projected points on the detector planes and the structural similarity index measure (SSIM) on the reconstructed volumes both before and after motion compensation. The ratio of RPE and SSIM after and before compensation depending on the cutoff frequency is plotted in Fig.~\ref{fig:overall}. The RPE ratio (left) is clearly below 1.0 for lower cutoff frequencies indicating a strong improvement of the RPE over the initial state. For the lowest investigated frequencies, the spline with 30 nodes yields slightly better results than the spline with 100 nodes. Both spline types perform on par for intermediate low frequencies around $f_c=0.02\omega$. The results obtained with 30 nodes become worse (increase in RPE ratio) when approaching the maximum node-dependent measurable frequency. The spline with 100 nodes shows the same behavior, but the increase in RPE happens at higher cutoff frequencies compared to the spline with fewer nodes. For completely unconstrained motion patterns, both splines lead to no, or only very minor improvement in RPE. 

The SSIM ratio plot (right) confirms these findings. In this case, higher values indicate an improved image quality after motion compensation. Again, we observe similar performance of both spline types for low frequencies up to the maximal frequency that can be recovered by 30 nodes. The SSIM ratio drops for the spline with 30 nodes, but stays high for the spline with 100 nodes until the respective critical frequency for 100 nodes is reached. Both splines do not yield a substantial improvement for unconstrained motion patterns. 

\begin{figure}
    \centering
    \includegraphics[width=\columnwidth]{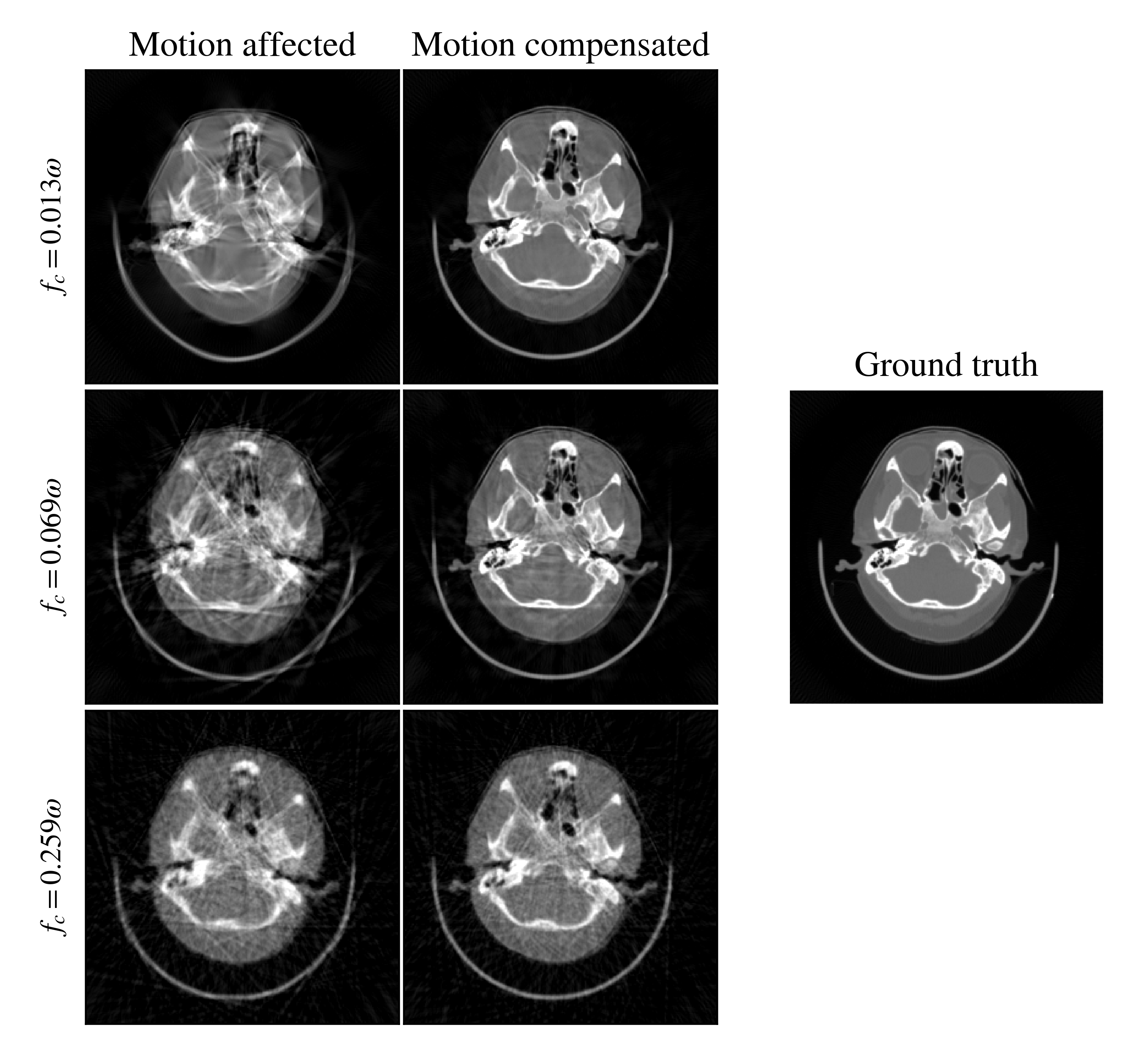}
    \caption{Reconstruction results for an axial slice of one example scan from three different cutoff frequencies for the perturbing motion pattern and a spline with 100 nodes for the estimated signal. The ground truth is identical for all three cases. Motion-compensated results appear best for the lowest frequency and worst for the highest frequency. }
    \label{fig:result_image}
\end{figure}
Fig.~\ref{fig:result_image} shows reconstructed results obtained from the same scan but with three different cutoff frequencies for the perturbing motion pattern. Looking at the motion-affected slices, we see that the frequency of the motion pattern has a strong influence on the appearance of the motion artifacts. Low frequencies lead to distortions of the shapes and misaligned edges, but some areas are still mostly homogeneous. High frequencies, in contrast, result in overall blurring of the reconstructed image and pronounced streak artifacts which affect all regions of the image. The motion-compensated results obtained with 100 nodes recover these artifacts well for the two lower frequencies and yield reconstructed images which resemble the ground truth closely. However, for $f_c=0.069\omega$ some inhomogeneities remain especially for soft tissue areas. The highest frequency in the figure $f_c=0.259\omega$ is above the critical frequency for 100 nodes according to the Nyquist-Shannon theorem. Therefore, the motion-compensated result sharpens the bone structures to some extent, but can not recover the blur and streaking artifacts caused by the high-frequency motion components.      

\section{Discussion}
The presented results confirm that the choice of motion model has a profound influence on the types of motion patterns which can be recovered by a certain motion compensation algorithm. For the investigated method, the frequencies that can robustly be recovered are limited by the sampling rate according to Nyquist-Shannon given the number of nodes of the respective spline. In the plot showing the RPE ratio, it can be observed that the performance of the motion compensation begins to decrease for frequencies slightly lower than the critical recoverable frequency. This confirms that the optimization in the investigated motion compensation algorithm is able to fit the spline-based motion model almost to its theoretical limit in terms of recoverable frequencies. For low-frequency motion patterns, the spline with fewer nodes leads to larger improvements in RPE, but these do not seem to translate into the reconstructed image as the SSIM ratio is similar for 30 and 100 nodes even for the lowest frequencies. On the other hand, a higher number of nodes increases the frequency range that can be recovered. As a result, for the investigated method, it seems reasonable to prefer a larger number of nodes if the maximal expected frequency in the motion signal can not reliably be estimated. 

We did not test the method with splines of more than 100 nodes. In the setting of this study with 360 projections in total, such a spline has a node for every $360 / 99 = 3.64$ acquired projection images. Considering existing research about the characteristics of head motion during a scanning procedure \cite{Wagner2003}, we believe that this setting is sufficient to cover even the high-frequency motion patterns that could occur in practice.

The network regressing the quality metric from an intermediate, motion-affected reconstruction in the investigated motion compensation algorithm was trained on slow motion patterns parameterized by splines with ten nodes. Interestingly, this does not seem to have a detrimental influence on the target function for higher-frequency motion patterns. We conclude that the quality regression network generalizes well to other frequencies than those present in the training data. 

This study is limited to rigid head motion and assumes that the motion occurs inter-frame, i.e., in between the projections and not during the acquisition of a single projection. Furthermore, we only test an explicit, spline-based motion model. A comparison to an implicit smoothness constraint that regularizes the target function is a valuable future extension of this work.   

\section{Conclusion}
In this study, we assess the influence of the smooth motion parameterization on the motion-compensated reconstruction depending on the frequency spectrum of the motion signal. Our findings indicate that, for the investigated motion compensation algorithm, an increased spline node count accommodates higher motion frequencies while maintaining robustness for low frequencies. This observation validates the algorithm's efficacy under a relaxed smoothness constraint. Ultimately, the choice of motion model has a substantial impact on the recoverable motion patterns and should be chosen with care in the context of the clinical scan protocol and application.     

\section*{Acknowledgment}
The research leading to these results has received funding from the European Research Council (ERC) under the European Union’s Horizon 2020 research and innovation program (ERC Grant No. 810316). The authors gratefully acknowledge the scientific support and HPC resources provided by the Erlangen National High Performance Computing Center (NHR@FAU) of the Friedrich-Alexander-Universit\"at Erlangen-N\"urnberg (FAU). The hardware is funded by the German Research Foundation (DFG).

\ifCLASSOPTIONcaptionsoff
  \newpage
\fi



\bibliographystyle{IEEEtran}
\bibliography{bibliography}
\end{document}